\documentclass{elsart3p}

\usepackage{amssymb}
\usepackage{epsfig}

\begin{document}
\begin{frontmatter}

\title{Trial wave functions for High-Pressure Metallic Hydrogen}

\author[AQ]{Carlo Pierleoni\corauthref{cor}},
\corauth[cor]{Corresponding author.}
\ead{carlo.pierleoni@aquila.infn.it}
\author[UCSB]{Kris T. Delaney},
\ead{kdelaney@mrl.ucsb.edu}
\author[UIUC]{Miguel A. Morales},
\ead{mmorale@uiuc.edu}
\author[UIUC]{David M. Ceperley},
\ead{ceperley@uiuc.edu}
\author[LPTL]{Markus Holzmann}
\ead{markus@lptmc.jussieu.fr}

\address[AQ]{INFM-CNR SOFT and Physics Department, University of L'Aquila,
L'Aquila, I-67010, Italy}

\address[UCSB]{Materials Research Laboratory, University of California,
Santa Barbara, CA 93106-5121, USA}

\address[UIUC]{Physics Department, University of Illinois at Urbana-Champaign
Urbana, Illinois 61801, USA}

\address[LPTL]{LPTMC,  Universit\'e Pierre et Marie Curie, 4 Place Jussieu,
75005 Paris, France, and
LPMMC, CNRS-UJF,  BP 166, 38042 Grenoble, France}

\begin{abstract}
Many body trial wave functions are the key ingredient for accurate Quantum Monte Carlo estimates of total electronic energies in many electron systems. In the Coupled Electron-Ion Monte Carlo method, the accuracy of the trial function must be conjugated with the efficiency of its evaluation. We report recent progress in trial wave functions for metallic hydrogen implemented in the Coupled Electron-Ion Monte Carlo method. We describe and characterize several types of trial functions of increasing complexity in the range of the coupling parameter $1.0 \leq r_s \leq1.55$. We report wave function comparisons for disordered protonic configurations and preliminary results for thermal averages.
\end{abstract}

\begin{keyword}
Quantum Monte Carlo; Many-body wave functions; \emph{Ab-initio} methods; High pressure hydrogen.
\PACS: 64.70.Ja, 61.20.Ja, 62.50.+p, 67.90.+z
\end{keyword}
\end{frontmatter}

\newcommand{\vect}[1]{{\bf {#1}}}
\section{Introduction}\label{aba:sec1}
Modern \emph{ab initio} simulation methods for systems of electrons and nuclei mostly rely on Density Function Theory (DFT) for computing the electronic forces acting on the nuclei, and on Molecular Dynamics (MD) techniques to follow the real-time
evolution of the nuclei. Despite recent progress, DFT suffers from well-known limitations\cite{Martin04,FoulkesMitasNeedsRajagopal01}. As a consequence, current \emph{ab initio} predictions of metallization
transitions at high pressures, or even the prediction of structural phase transitions, are often only qualitative. Hydrogen is an extreme case\cite{MaksimivShilov99,StadeleMartin00,JohnsonAschroft00}, but even in silicon, the diamond/$\beta$-tin transition pressure and the melting temperature are seriously underestimated\cite{AlfeGillanTowlerNeeds04}.

An alternative route to the ground-state properties of a many-electrons system is the Quantum Monte Carlo method (QMC)\cite{HammondLesterReynolds94,FoulkesMitasNeedsRajagopal01}. In QMC, a many-body trial wave function for the electrons is assumed and the electronic properties are computed by Monte Carlo methods. Bosonic details of the trial wave functions are automatically optimized by projecting the trial state onto the ground state with the same nodal structure. Hence for fermions, QMC is a variational method with respect to the nodes of the trial wave function and a systematic, often unknown, error remains\cite{HammondLesterReynolds94,FoulkesMitasNeedsRajagopal01}.
Over the years, the level of accuracy of the fixed-node approximation has been improved\cite{PanoffCarlson89,KwonCeperleyMartin94,HolzmannCeperleyPierleoniEsler03,RiosMaDrummondTowlerNeeds06}
such that, in most cases, fixed-node QMC methods have proven to be more accurate than DFT-based methods, on one side, and less computationally demanding than correlated quantum-chemistry strategies (such as coupled cluster method)\cite{FoulkesMitasNeedsRajagopal01} on the other side.
Recently we have developed an ab-initio simulation method, the Coupled Electron-Ion Monte Carlo (CEIMC) method, based entirely on Monte Carlo algorithms, both for solving the electronic problem and for sampling the ionic configuration space in the Born-Oppenheimer approximation\cite{PierleoniCeperley05}.
A Metropolis Monte Carlo simulation of the ionic degrees of freedom (represented either by classical point particles or by path integrals) at fixed temperature is performed based on the electronic energies computed during independent ground state Quantum Monte Carlo calculations.
Application of CEIMC has been limited so far to high pressure hydrogen for several reasons:
a) hydrogen is the simplest element of the periodic table, and the easiest to cope with since the absence of the additional separation of energy scales between core and valence electrons as in heavier elements;
b) it is an important element since most of the matter in the universe consists of hydrogen;
c) its phase diagram at high pressure in the interesting region where the metallization occurs is still largely unknown because present experiments are not able to reach the relevant pressures.
We have investigated the very high pressure regime where all molecules are dissociated and the system is a plasma of fully ionized protons and electrons\cite{PierleoniCeperleyHolzmann04}, and we have studied the pressure-induced molecular dissociation transition in the liquid phase\cite{DelaneyPierleoniCeperley06}. In both studies the CEIMC results were not
in agreement with previous Car-Parrinello Molecular Dynamics (CPMD) calculations\cite{KohanoffHansen96,Scandolo03}.
While we have evidence now that the discrepancy in the fully ionized case is removed by taking a more accurate trial wave function in the QMC, in the second study more accurate CEIMC calculations predict a continuous molecular dissociation with increasing pressure at variance with CPMD where a first order molecular dissociation transition was observed by increasing pressure at constant temperature\cite{Scandolo03}. More recently, using constant volume Born-Oppenheimer Molecular Dynamics rather then constant pressure CPMD, a continuous dissociation transition has been reported from DFT-GGA studies \cite{VorbergerTamblynMilitzerBonev07}.

In the present paper we discuss in some details the various trial wave functions we have implemented for hydrogen and we compare their accuracy at various densities. We will not review the details of the CEIMC method which have been described at length in reference \cite{PierleoniCeperley05}. We just mention that in metals huge finite size effects, caused by the discrete nature of the reciprocal space, can be alleviated by averaging electronic properties over the overall undetermined phase of the many-body wave function (Twist Average Boundary Conditions). Results reported in this work are obtained by this method (see also ref.\cite{PierleoniDelaneyMoralesCeperleyHolzmann07} for recent development).
Section \ref{sec2} will be devoted to describing the different trial wave functions and some details on their efficient implementation. In Section \ref{sec3} we will report numerical comparisons among the various trial functions. Finally, in Section \ref{sec4} we collect our conclusions and perspectives.

\section{Trial Wave Functions for Hydrogen}\label{sec2}
The trial wave functions we have adopted for hydrogen are of the simple Slater-Jastrow form. We have considered spin unpolarized hydrogen only. For each spin state, a single determinant of one electron orbitals $\phi_k(\vect{r})$ is used to account for the fermionic symmetry of the many-body wave function. A Jastrow factor $e^{-U}$ is then added to account for (at least) two body correlations directly into the trial wave function
\begin{equation}
\Psi_T(R,S)=D_{\uparrow}D_{\downarrow}e^{-U}
\end{equation}
with
\begin{equation}
D_\xi = \left|
\begin{array}{cccc}
\phi_1\left(\vect{r}_1\right) & \phi_1\left(\vect{r}_2\right) & \ldots & \phi_1\left(\vect{r}_{N/2}\right)\\
\phi_2\left(\vect{r}_1\right) & \phi_2\left(\vect{r}_2\right) & \ldots & \phi_2\left(\vect{r}_{N/2}\right)\\
  \vdots & & \ddots\\
\phi_{N/2}\left(\vect{r}_1\right) & \phi_{N/2}\left(\vect{r}_2\right) & \ldots & \phi_{N/2}\left(\vect{r}_{N/2}\right)\\
\end{array}
\right|
\end{equation}
and $U=\sum_{i<j} u_{ij}(r_{ij})$. An accurate and parameter free Jastrow factor can be obtained within the RPA $u^{RPA}_{ij}(r)~(i,j=e,p)$ \cite{Gaskell61}. This simple form satisfies the correct cusp conditions at short particle separations and the right plasmon behavior (screening) at large distances. It was shown\cite{CeperleyAlder87} to provide good energies for hydrogen even at intermediate densities if supplemented by gaussian functions $\tilde{u}_{ij}(r)=u^{RPA}_{ij}(r)-\alpha_{ij}e^{-{r^{2}}/{w_{ij}^{2}}}$, with the variational parameters $\alpha_{ij}, w_{ij}$. The additional terms preserve the short- and long-distance behavior of the RPA function and correct for possible inaccuracies at intermediate distances. However, they introduce four variational parameters to be numerically optimized, namely $\alpha_{ee}, w_{ee}, \alpha_{ep}, w_{ep}$. We have adopted the RPA form of the two body correlation in all different trial function discussed below.

In Ref. \cite{HolzmannCeperleyPierleoniEsler03} we have described a procedure to iteratively improve any initial trial function based on the Feynman-Kac formula. If a determinant of single electron orbitals is assumed as an initial ansatz, the first iteration generates a bosonic (symmetric) two-body correlation function (Jastrow) while the next iteration naturally provides the backflow transformation of the orbitals and a three-body bosonic correlation term. Unfortunately, this is a formal theory which cannot provide, in general, analytical expressions for the various terms. Nonetheless the general structure is illuminating in searching for improvements.

In the first implementation of CEIMC\cite{DewingCeperley02,CeperleyDewingPierleoni02}m single electron orbitals consisting of a linear combination of few optimizable guassian orbitals centered on each molecule was used in order to simulate the insulating phases of molecular hydrogen. Optimization of the variational parameters, in number proportional to the number of electrons in the system, needed to be performed at each ionic configuration and was a major bottleneck for the efficiency of the method. Subsequently, we have developed trial functions with a very limited number of variational parameters (even zero when possible) and therefore reduced the complexity of the optimization step in the electronic calculation (or to reduce to a linear optimization in the case of DFT orbitals).

\subsection{The Metallic Wave Function}
At a very high pressure, beyond metallization and molecular dissociation, the electron liquid is a good Fermi liquid and correlation effects, with protons and among electrons, can be treated as perturbations. In this case it is natural to assume a determinant of free electron states (plane waves) as an initial ansatz. As stated above, the first iteration provides the two-body Jastrow factor for which we adopted the modified RPA form discussed above. Although this form of the trial function has been extensively used in the early QMC study of hydrogen\cite{CeperleyAlder87}, it is worthwhile to emphasize that the proton coordinates do not appear in the fermionic part of the trial function, that is in its nodal structure, which is ultimately the limit of the accuracy of QMC. This limitation is overcome at the next iteration in the Feynman-Kac procedure, which suggests the backflow transformation of the orbitals and a three-body correlation factor.
The form of the backflow transformation is
\begin{equation}
  \vect{x}_i = \vect{r}_i + \sum_j \eta_{ij}\left(\left|r_{ij}\right|\right)\vect{r}_{ij},
\end{equation}
where $\eta_{\alpha\beta}$ are the electron-electron and electron-proton backflow functions that must be parameterized. When the single body-orbitals in the determinants are expressed in terms of the quasi-particle coordinates $\vect{x}_i$, the nodal surfaces of the trial wave function become explicitly dependent on the proton positions, a crucial characteristic for inhomogeneous electron systems which will provide a more accurate energy even at the RQMC level. Similar to the case of the homogeneous electron gas\cite{KwonCeperleyMartin94}, the backflow and three-body functions were at first parametrized as gaussians\cite{CeperleyDewingPierleoni02}. This trial function has a total of 10 free parameters to be variationally optimized and has been used in a first CEIMC study of the melting transition of the proton crystal in hydrogen at $r_s=1$ \cite{CeperleyDewingPierleoni02}. Subsequently, we were able to derive approximate analytical expressions for the backflow and the three-body functions, as well as for the two-body correlation factor, in the Bohm-Pines collective coordinates approach \cite{HolzmannCeperleyPierleoniEsler03}. This form is particularly suitable for the CEIMC because it is parameter-free. At the same time, it provides comparable accuracy to the numerically optimized wave function, both in the crystal configuration and for disordered protons. Explicit forms of the various terms can be found in the appendix of Ref.\cite{HolzmannCeperleyPierleoniEsler03}. With this kind of wave function we have investigated the melting at three densities ($r_s=0.8, 1.0, 1.2$) including quantum effects for protons\cite{PierleoniCeperleyHolzmann04,PierleoniCeperley05}.

\subsection{Band-structure-based Wave Functions (IPP/LDA)}\label{secbands}
The metallic wave function is expected to provide an accurate description of the electronic ground state at high density, well beyond molecular dissociation and metallization. On the other hand we expect it to be a poor representation of the true ground
state at lower densities near the molecular dissociation. At lower densities, molecules appear and plane-wave single-electron orbitals (although in terms of backflow coordinates) are certainly not a good representation. Natoli \emph{et al.}\cite{NatoliMartinCeperley93,NatoliMartinCeperley95} have previously used a Slater determinant of Kohn-Sham self consistent orbitals to study the solid phases of atomic hydrogen at $r_s=1.31$ and $T=0$, and of molecular hydrogen at lower densities. They have found a typical energy gain of 0.5eV/electron by replacing the plane-wave with the self consistent orbitals in the Slater determinant. Here we have implemented similar ideas.

The single-particle orbitals, $\left\{\phi_n\right\}$, that comprise the Slater determinant are computed on-the-fly during the
CEIMC calculation as the eigenstates of some single-particle Hamiltonian
\begin{equation}
\hat{h}\phi_n\left(\vect{r}\right) = \varepsilon_n\phi_n\left(\vect{r}\right),
\end{equation}
where the $N/2$ orbitals with the lowest eigenvalue, $\varepsilon_n$, are selected to fill the determinants. The single-particle Hamiltonian describes electron-nuclear interactions and approximates electron-electron interactions through
an effective potential. In Hartree atomic units
\begin{equation}
\hat{h} = -\frac{1}{2}\nabla^2 + V_\mathrm{eff}\left(\vect{r};S\right).
\end{equation}
The effective potential, $V_\mathrm{eff}\left(\vect{r};S\right)$, depends on the nuclear coordinates, $S$, as parameters and is, by construction, invariant under translation of the electron coordinate from one simulation cell to another:
\begin{equation}
V_\mathrm{eff}\left(\vect{r} + \vect{L};S\right) = V_\mathrm{eff}\left(\vect{r};S\right).
\end{equation}
The many-body Slater-Jastrow wave function should obey twisted boundary conditions\cite{LinZongCeperley01}.
The effect is that translating any single-electron coordinate by a lattice vector twists the overall phase of the wave function
\begin{equation}
\Psi\left(\vect{r}_1,\dots,\vect{r}_i+\vect{L},\dots,\vect{r}_N\right) = e^{i\vect{\theta}}\Psi\left(\vect{r}_1,\dots,\vect{r}_i,\dots,\vect{r}_N\right)
\end{equation}
which is achieved by enforcing Bloch's theorem on the single-particle orbitals for the first-Brillioun zone wavevector corresponding to the twist angle
\begin{equation}
\phi_{n\vect{k}}\left(\vect{r}\right) = U_{n\vect{k}}\left(\vect{r}\right)e^{i\vect{k}.\vect{r}}; \vect{k} \in \mathrm{1BZ}.
\end{equation}
The Bloch functions, $U_{n\vect{k}}$, are then cell-periodic functions, which are particularly well-represented in a plane-wave basis set, with basis functions of the form
\begin{equation}
B_\vect{G}\left(\vect{r}\right) = \frac{1}{\sqrt{V}}e^{i\vect{G}.\vect{r}}.
\end{equation}
The basis functions are orthogonal in $\vect{G}$ and normalized to $1$ through the simulation cell of volume $V$. The $G$ vectors are vectors of the reciprocal-space lattice, defined such that $\vect{G}.\vect{L} = 2m\pi; m\in \mathbb{Z}$. The quality of the plane-wave basis is controlled by a single parameter, an energy cutoff, which determines that largest wavevector that is present in the set:
\begin{equation}
\frac{1}{2}G^2 \leq E_\mathrm{cut}.
\end{equation}

In the plane-wave basis, the orbitals can be expressed as
\begin{equation}
\phi_{n\vect{k}}\left(\vect{r}\right) = \frac{1}{\sqrt{V}}\sum_\vect{G} C_{n\vect{k}\vect{G}}e^{i\left(\vect{k}+\vect{G}\right).\vect{r}}.
\end{equation}
The problem of filling the Slater determinant reduces to a problem of finding the plane-wave coefficients, $C_{n\vect{k}\vect{G}}$, of the Bloch states and evaluating the resulting determinant for the QMC electronic configuration. The plane-wave orbital coefficients are found by solving the eigenvalue problem of the single-particle Hamiltonian
\begin{equation}
\sum_{\vect{G}^\prime} \left<B_\vect{G} \left| \hat{h}_\vect{k} \right| B_{\vect{G}^\prime} \right> C_{n\vect{k}\vect{G}^\prime} = \varepsilon_{n\vect{k}} C_{n\vect{k}\vect{G}},
\end{equation}
where
\begin{eqnarray}
\left<B_\vect{G} \left| \hat{h}_\vect{k} \right| B_{\vect{G}^\prime} \right> &=& -\frac{1}{2}\left|\vect{k}+\vect{G}\right|^2\delta_{\vect{G},\vect{G}^\prime} \\
&+& \frac{1}{V}\int_\mathrm{cell}V_\mathrm{eff}\left(\vect{r};S\right)e^{i\left(\vect{G}-\vect{G}^\prime\right).\vect{r}}d^3\vect{r} \nonumber
\label{eqn:ftham}
\end{eqnarray}
From this Fourier transform, it is clear that if the simulation cell is centrosymmetric then the potential is real and inversion symmetric in $\vect{G}-\vect{G}^\prime$. However, this and other symmetry considerations, such as the existence of
point-group operations that leave the single-particle Hamiltonian invariant, are generally not relevant in a CEIMC simulation in which the nuclear coordinates are typically not ordered. One relevant symmetry, time-reversal symmetry for $\vect{k}\rightarrow-\vect{k}$, implies that only half of the Brillioun zone must be explicitly sampled. Further, the eigenvectors corresponding to $\vect{k}=0$ are real.

Note that the effective potential in reciprocal space depends on $\vect{G}-\vect{G}^\prime$ only, and can therefore be stored as a one-dimensional vector of size $8N_G$ in place of an array of size $N_G^2$. This property is due to the locality of the potential in real space.
One method of solving the eigenvalue problem would be to use standard numerical matrix diagonalization techniques. However, such approaches are much too slow when the number of basis functions significantly exceeds the number of required eigenstates, as in the case of a plane-wave basis. Instead, we use an iterative, conjugate-gradient band-by-band minimization scheme\cite{PayneReview92} that is particularly effecient for these classes of problems. The method employs the
variational principle to minimize residuals and the Gram-Schmidt scheme to preserve orthogonalization of the eigenstates.
In our studies, we typically use one of two choices for the effective potential:
\begin{itemize}
\item $V_\mathrm{eff} = V_{e-n}$, the bare electron-nuclear interaction. We refer to this method as the independent particle potential (IPP) because electron-electron interactions and nuclear screening are absent from the orbitals.
\item $V_\mathrm{eff} = V_\mathrm{LDA}$, the Kohn-Sham effective potential within the local density approximation (LDA).
\end{itemize}
Further complexities may be introduced to all of these wave functions through backflow transformations and the Jastrow factor.

\subsection{IPP and LDA Orbitals}
For hydrogen, the IPP orbitals are generated with $V_\mathrm{eff}$ from the bare electron-proton Coulomb potential such that the potential term from Equation \ref{eqn:ftham} becomes
\begin{equation}
  V_{ep}\left(\vect{G}-\vect{G}^\prime\right) = \frac{1}{V}S_{\vect{G}-\vect{G}^\prime} v_{c}\left(\left|\vect{G}-\vect{G}^\prime\right|\right),
\end{equation}
where $S_\vect{G}$ is the simulation-cell structure factor
\begin{equation}
S_{\vect{G}} = \sum_I e^{-i\vect{G}.\vect{R}_I},
\end{equation}
with $R_I$ a proton coordinate, and $v_{c}$ is the Fourier transform of the Coulomb potential
\begin{equation}
v_{c}\left(G\right) = -\frac{4\pi}{G^2}.
\end{equation}
The diverging $\vect{G}=\vect{G}^\prime$ components are equated to zero, which corresponds to shifting the arbitrary zero of electrostatic potential.

DFT-LDA orbitals are generated from the Kohn-Sham effective potential with the local density approximation. The effective potential becomes a sum of three terms
\begin{equation}
V_\mathrm{eff}\left(\vect{r}\right) = V_{ep}\left(\vect{r};S\right) + V_\mathrm{Ha}\left(\vect{r}\right) + V_{xc}^\mathrm{LDA}\left(\vect{r}\right).
\end{equation}
Both the Hartree potential and the exchange-correlation potential depend on the electron density
\begin{equation}
n\left(\vect{r}\right) = \sum_n^\mathrm{occ}\left|\phi_n\left(\vect{r}\right)\right|,
\end{equation}
which in turn depends on the eigenfunctions of the Hamiltonian. These potentials must therefore be determined self-consistently, through an iterative procedure, such that the Hamiltonian yields orbitals corresponding to a density equal to that used to generate the potentials. The local density approximation that we use is the Perdew-Zunger\cite{PerdewZunger81}
parameterization of Ceperley-Alder\cite{CeperleyAlder80} electron-gas data.

\begin{figure}[h]
  \begin{center}
    \resizebox{1.0\columnwidth}{!}{\includegraphics{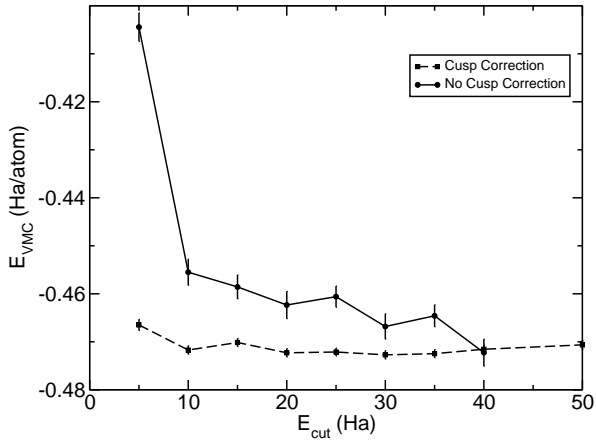}}
    \caption{VMC total energy for a fixed proton configuration and a Slater-Jastrow
    trial wave function with IPP orbitals. The VMC energy converges slowly with
    respect to the orbital plane-wave basis cutoff, but this is greatly improved
    by removing the inaccurate $ep$ cusp and correcting in real space
    with an analytic Jastrow function.}
    \label{fig:cuspremove}
  \end{center}
\end{figure}

For hydrogen, both the LDA and IPP wave functions are eigenstates of a Hamiltonian which contains a bare Coulomb interaction between electrons and protons. The singularity in the potential results in a derivative cusp in $\phi\left(\vect{r}\right)$ when $\vect{r}=\vect{R}_I$. The presence of this cusp with orbitals that properly satisfy Kato's cusp condition is important
for obtaining good energies or short projection times for QMC algorithms. The electron-proton cusp condition is:
\begin{equation}
  \left.\frac{\partial \phi\left(\vect{r}\right)}{\partial \vect{r}}\right|_{\vect{r}=\vect{R}_I} = -\phi\left(\vect{r}=\vect{R}_I\right)
\end{equation}
Representing this cusp in reciprocal space is challenging due to the slow algebraic decay of $G^{-4}$. For this reason, we implement a cusp-removal method by dividing the orbitals by a function that satisfies the cusp condition exactly (in this case,
the RPA $ep$ Jastrow function discussed earlier),
\begin{equation}
  \tilde{\phi}_{n\vect{k}}\left(\vect{r}\right) = \frac{\phi_{n\vect{k}}\left(\vect{r}\right)}{e^{-\sum_I u_{ep}^\mathrm{RPA}\left(\vect{r}-\vect{R}_I\right)}},
\end{equation}
before reverse Fourier transforming to retrieve the $C_{n\vect{k}\vect{G}}$ coefficients used to build the Slater determinant.
The proper electron-proton cusp is later reintroduced exactly in real space using the same RPA Jastrow function. This procedure greatly enhances the convergence of the Slater-Jastrow wave function with respect to the size of the plane-wave basis set used to represent the orbitals, as demonstrated by Figure \ref{fig:cuspremove}.

\subsection{Backflow Transformations of IPP/LDA Orbitals}
Similarly to the metallic wave function case, we can think of applying the Feynman-Kac iteration procedure to generate a backflow transformation even for IPP/LDA orbitals. This transformation introduces correlation effects into the Slater determinant part of a Slater-Jastrow wave function, with the advantage that modifications of the nodal surface are possible and the energy is improved.
The specific form for the transformation is unknown, but as a first ansatz we can use the same analytical expressions we have developed in the metallic case. As we will show in the next section, backflow is found to improve the total energy and the variance of the total energy, both indicating that it is a better representation of the ground state of the system. Only the electron-electron backflow is applied, while the electron-proton term is found to reduce the quality of the trial function.

\section{Comparisons of Wave Functions}\label{sec3}
\subsection{Fixed protonic configurations}
In this section we consider fixed protonic configurations and we compare the quality of the various wave functions at several densities. We consider first protons in crystal structures at $r_s=1.0$ and $r_s=1.4$. Results for BCC hydrogen at $r_s=1.31$ obtained from various improvements of the metallic wave function are reported in Table III of Ref. \cite{HolzmannCeperleyPierleoniEsler03} where they are also compared to the results obtained with self-consistent Kohn-Sham orbitals\cite{NatoliMartinCeperley93}. There we have shown that the quality of the analytical form of the metallic wave function is superior to its numerically optimized version and comparable to that of the LDA orbitals for hydrogen in the BCC structure and for various system sizes. Our present implementation of LDA orbitals provides results in agreement with previous estimates\cite{Delaney}.
\begin{table}
\caption{Energy and variance of hydrogen in various structures with different trial functions at $r_s=1.0$. All results are obtained averaging over a 6x6x6 fixed grid of twist angles. $E_v$ and $\sigma^2_v$ represent, respectively, energy and variance at the variational level while $E_{r}$ and $\sigma^2_{r}$ are the energy and the mixed estimator for the variance obtained with RQMC. Units are Hartree/atom.}
\resizebox{1.00\columnwidth}{!}{\begin{tabular}{|c|c|cccc|}
\hline
&WFS&$E_v$&$\sigma^2_v$&$E_{r}$&$\sigma^2_{r}$\\
\hline
               &Met   & -0.36931(1)& 0.0279(2) & -0.3721(1) & 0.0182(7)\\
BCC            &IPP   &             &           & -0.3681(3) & 0.0765(3)    \\
($N_p=54$)     &LDA   &             &           & -0.3681(2) & 0.0765(2)    \\
               &IPP+BF&             &           & -0.3705(1) & 0.0359(1)   \\
               &LDA+BF&-0.36805(2)&0.04636(4)& -0.3705(1) & 0.0357(1)  \\
               \hline
               &Met   &             &           & -0.3792(1) & 0.01543(4) \\
FCC            &IPP   &             &           & -0.3756(2) & 0.0756(2)   \\
($N_p=32$)     &LDA   &             &           & -0.3757(2) & 0.0753(2)  \\
               &IPP+BF&             &           & -0.3779(1) & 0.03550(9) \\
               &LDA+BF&             &           & -0.3779(1) & 0.0351(1)  \\
               \hline
               &Met   &             &           & -0.3477(2) & 0.0268(1)   \\
DIAM           &IPP   &             &           & -0.3621(2) & 0.0830(4)   \\
($N_p=64$)     &LDA   &             &           & -0.3613(2) & 0.0823(3)   \\
               &IPP+BF&             &           & -0.3637(1) & 0.0404(1)   \\
               &LDA+BF&             &           & -0.3635(1) & 0.0406(1)  \\
               \hline
               &Met   & -0.41060(4) & 0.02136(4)& -0.41368(6)& 0.01032(2) \\
               &IPP   & -0.40198(8) & 0.0863(1) & -0.4094(1) & 0.04342(8) \\
DIAM           &LDA   & -0.40206(8) & 0.0865(1) & -0.4098(1) & 0.04356(6)  \\
($N_p=8$)      &IPP+BF& -0.40632(6) & 0.04958(8)& -0.41070(6)& 0.02382(4) \\
               &LDA+BF& -0.40638(6) & 0.04958(8)& -0.4107(1) & 0.02382(6)  \\
\hline
\end{tabular}}\label{table1}
\end{table}

\begin{table}
\caption{Energy and variance of hydrogen in various structures with different trial functions at $r_s=1.4$. All results are obtained averaging over a 6x6x6 fixed grid of twist angles. $E_v$ and $\sigma^2_v$ represent, respectively, energy and variance at the variational level while $E_{r}$ and $\sigma^2_{r}$ are the energy and the mixed estimator for the variance obtained with RQMC. Units are Hartree/atom.}
\resizebox{1.00\columnwidth}{!}{\begin{tabular}{|c|c|cccc|}
\hline
&WFS&$E_v$&$\sigma^2_v$&$E_{r}$&$\sigma^2_{r}$\\
\hline
               		&Met   	& -0.5203(2)	 & 0.036(2) & -0.5224(1) & 0.01008(4) \\
BCC          	&IPP     	& -0.5139(1) 	& 0.0887(2)& -0.5209(2) & 0.0284(1)  \\
($N_p=54$)     &LDA   	& -0.5145(1) 	& 0.0869(3)& -0.5210(2) & 0.0286(1)  \\
               		&IPP+BF	&            		&          & -0.5228(1) & 0.01413(7) \\
\hline
               		&Met   &            &          & -0.5272(1) & 0.00872(3) \\
FCC            	&IPP     & -0.5210(1) & 0.0835(3)& -0.5256(1) & 0.0276(1)  \\
($N_p=32$)     &LDA   & -0.5212(1) & 0.0828(3)& -0.5259(1) & 0.02724(9) \\
               		&IPP+BF&            &          & -0.5280(1) & 0.01352(5) \\
\hline
               		&Met   &            &          & -0.5168(1) & 0.01656(8) \\
DIAM           	&IPP  & -0.5189(2) & 0.1027(8)& -0.5321(5) & 0.0339(3)  \\
($N_p=64$)     &LDA   &            &          & -0.5323(1) & 0.0331(2)  \\
               		&IPP+BF&            &          & -0.5346(1) & 0.01740(7) \\
\hline
\end{tabular}}\label{table2}
\end{table}
In Tables \ref{table1} and \ref{table2} we report QMC energies for hydrogen in several crystal structures and for various system sizes at two densities corresponding to $r_s=1$ and $r_s=1.4$ respectively. A complete study of the size dependence and the relative stability of those structure is not our concern here and will be reported elsewhere\cite{Delaney}. We observe that LDA always provides a small or negligible improvement over IPP, while IPP is significantly cheaper through the lack of the self-consistent requirement. Comparing various structures and system sizes, we observe that the best wave function depends on the structure: for BCC, FCC structures and the diamond structure with N=8, the metallic wave function is superior to the others. The opposite is true for the diamond structure with N=64, where IPP and LDA provide lower energies at all densities. We also observe that the ordering of the wave functions does not appear to depend on density, at least in the limited range investigated. Note that $r_s=1.31$ corresponds to the density predicted by ground state QMC calculations\cite{CeperleyAlder87} for the molecular dissociation to occur. Another important test is the effect of backflow (BF) on the band orbitals. We see that both variation and reptation energies are slightly improved and the variational variance is halved by the backflow, which means a net improvement of the trial function and the need of a shorter projection in imaginary time to reach the ground state. The high level of accuracy observed for the metallic wave function in crystal structures induced us to perform a detailed study of liquid hydrogen at finite temperature\cite{PierleoniCeperleyHolzmann04}.

Next, we consider how the various wave functions perform on disordered protonic configurations representative of atomic hydrogen in the liquid state in the range of densities corresponding to the interval $r_s\in[1.0,1.55]$. As before, all results reported here are averaged over a 6x6x6 fixed grid in the twist space. At $r_s=1$ we compare the Metallic wave function with the LDABF wave function, while at lower densities we report data for the IPP, LDA and LDABF wave functions.
\vskip 0.5cm
\begin{figure}[h]
\begin{center}
\resizebox{1.0\columnwidth}{!}{\includegraphics{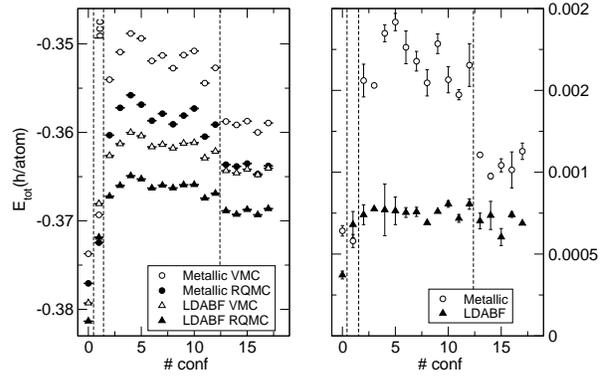}}
\end{center}
\caption{Total energy (left panel) and quality parameter (right panel) for a number of static proton configurations as obtained with the metallic and the LDABF trial functions at $r_s=1.0$. TABC with a 6x6x6 fixed grid in the twist space is performed. Energies are in h/atom. The quality parameter is defined in the text.}
\label{fig:rs1static}
\end{figure}
In the left panel of Figure \ref{fig:rs1static} we display VMC and fully converged RQMC energies for 18 protonic
configurations at $r_s=1.0$ obtained with the metalllic and the LDABF wave functions. Configuration 0 is a
32 protons warm crystal (FCC) near melting, configuration 1 is the perfect BCC crystal with 54 protons,
configurations 2 to 12 are statistically independent configurations of 54 protons obtained
during a CEIMC run at T=2000K performed with the LDABF trial function, while the remaining 5
configurations have been obtained during a CEIMC run at the same temperature performed with the
metallic trial function.
Several interesting facts can be inferred from this figure. With the noticeable exception of the
perfect BCC crystal, energies from LDABF wave function are always lower than energies from the metallic wave function.
In particular, the fully converged RQMC energies from the metallic wave functions are above the
VMC energies from LDABF trial function. This implies that changing the form of the nodes provides more energy than fully projecting the initial state.
Why the excellent quality of the metallic wave function observed in perfect crystals is deteriorated by disorder remains unclear to us,
but as a matter of fact, it appears that LDA nodes supplemented by e-e backflow perform much better both in the liquid state and in the crystal state with thermal fluctuations. Another interesting observation concerns the dispersion of the energies over a set of configurations. The dispersion of the energy, measured by its standard deviation with respect to the mean value, is a measure of the roughness of the BO energy surface, the key quantity in determining the structure and the thermal properties of hydrogen.
The dispersions over the first 11 liquid configurations (from conf. 2 to conf. 13 generated during a run with the LDABF trial function) is reported in Table \ref{table3}. Dispersion of the RQMC energies with the metallic trial function is roughly twice as large than the dispersion from the LDABF trial function which indicates as the BO surface with LDABF is smoother than the other, providing a less structured liquid state and a lower melting temperature of the proton crystal. Moreover, it is interesting to compare the dispersion of the VMC and the RQMC energies for a given trial function and a given set of configurations. For the same set of liquid configurations and for the LDABF wave function, the VMC and RQMC dispersions differ by 0.09mH/at corresponding to roughly 30K, a tiny difference in the roughness of the BO energy surface. This suggest that at $r_s=1.0$ the VMC BO energy surface is likely to be accurate enough.
\begin{figure}[t]
\begin{center}
\psfig{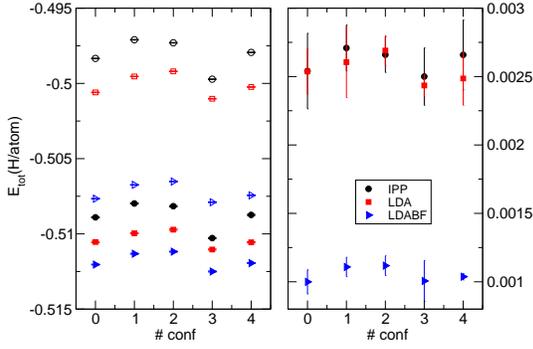}
\end{center}
\caption{Total energy (left panel) and quality parameter (right panel) for a number of static proton configurations as obtained with the metallic and the LDABF trial functions at $r_s=1.31$. TABC with a 6x6x6 fixed grid in the twist space is performed. Energies are in h/atom. In the right panel open symbol represent VMC energies for IPP (circles), LDA (squares) and LDABF (triangles), respectively. RQMC energies for the same trial functions are represented by closed symbols.}
\label{fig:rs1.31static}
\end{figure}
\begin{figure}[t]
\begin{center}
\psfig{file=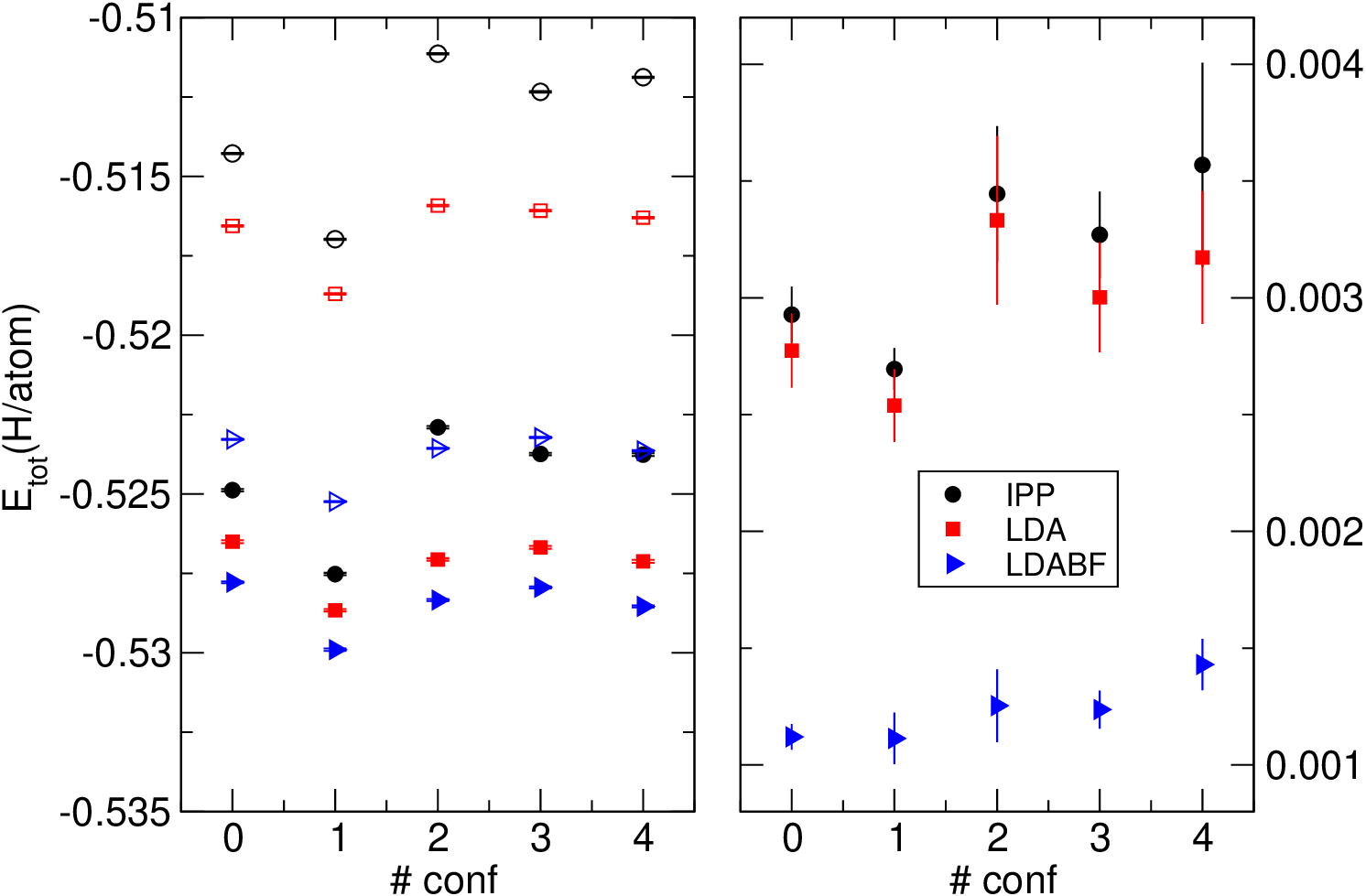,width=7cm}
\end{center}
\caption{Total energy (left panel) and quality parameter (right panel) for a number of static proton configurations as obtained with the metallic and the LDABF trial functions at $r_s=1.40$. TABC with a 6x6x6 fixed grid in the twist space is performed. Energies are in h/atom. In the right panel open symbol represent VMC energies for IPP (circles), LDA (squares) and LDABF (triangles), respectively. RQMC energies for the same trial functions are represented by closed symbols.}
\label{fig:rs1.4static}
\end{figure}
\begin{figure}[t]
\begin{center}
\psfig{file=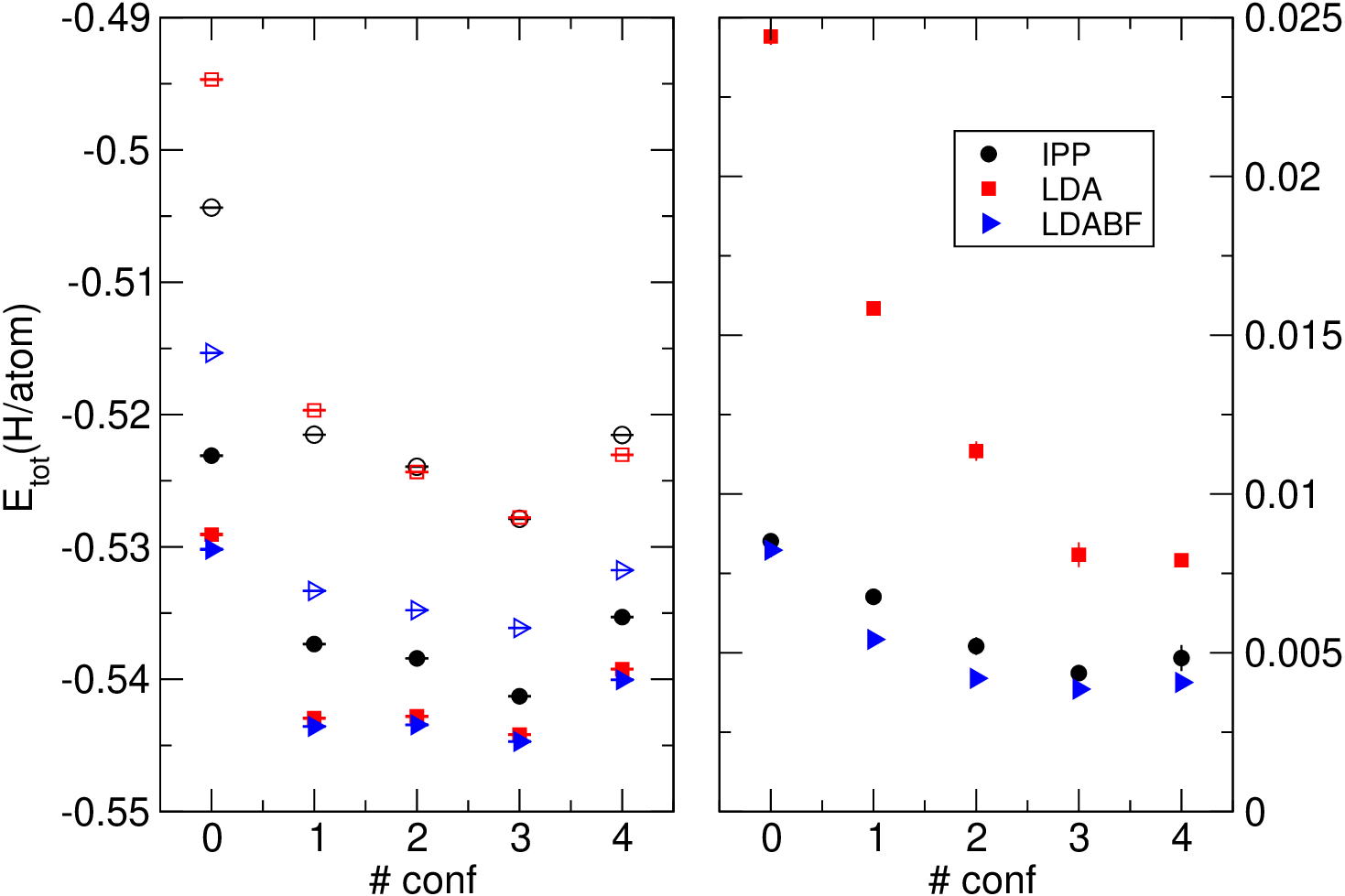,width=7cm}
\end{center}
\caption{Total energy (left panel) and quality parameter (right panel) for a number of static proton configurations as obtained with the metallic and the LDABF trial functions at $r_s=1.55$. TABC with a 6x6x6 fixed grid in the twist space is performed. Energies are in h/atom. In the right panel open symbol represent VMC energies for IPP (circles), LDA (squares) and LDABF (triangles), respectively. RQMC energies for the same trial functions are represented by closed symbols.}
\label{fig:rs1.55static}
\end{figure}
\begin{table}
\begin{center}
\caption{Dispersion of the energy over representative static configurations at various densities. All results are obtained averaging over a 6x6x6 fixed grid of twist angles. Units are mHartree/atom.}
\resizebox{0.80\columnwidth}{!}{
{\begin{tabular}{|c|cccc|}
\hline
		&\multicolumn{4}{c}{VMC}\\
\hline
$r_s$	&Met		&IPP		&LDA	&LDABF\\
\hline
1.0          	& 1.666(1)	&   --    	&   --    	 & 0.820(1) \\
1.31	        	&   --    	 	& 0.93227(2)& 0.66897(2) & 0.53286(1)  \\
1.40		&   -- 		& 2.10255(4)& 1.01702(2) & 0.74335(1)  \\
1.55         	&   -- 		& 2.59550(5)& 2.9005(1) & 1.62655(3) \\
\hline
		&\multicolumn{4}{c}{RQMC}\\
\hline
$r_s$	&Met		&IPP		&LDA	&LDABF\\
\hline
1.0          	& 1.380(3)	&   --    	&   --    	& 0.730(2) \\
1.31	        	&   --    	 	& 0.81059(3)& 0.46996(2) & 0.48619(1)  \\
1.40		&   --    	 	& 1.60822(5)& 0.76393(3) & 0.75089(1)  \\
1.55         	&   --    	 	& 2.15821(6)& 1.8410(1) & 1.74226(5) \\\hline
\end{tabular}}}\label{table3}
\end{center}
\end{table}
In the right panel of Figure \ref{fig:rs1static} we report for the same static configurations the quality parameter for the two wave functions considered. The quality parameter of a trial function is defined as the negative logarithm of the overlap of the trial state onto its fully projected state. It is easy to prove that it reduces to the integral over positive imaginary time  of the difference between the energy and its extrapolation at infinite time. The smaller the quality parameter the better the trial function is. We observe that, with the exception of the BCC crystal, the metallic wave function has larger values, which means that it is less accurate than the LDABF wave function. Note also, how the quality of the LDABF wave function is uniform (at fixed number of particles)
through the perfect crystal and the disordered configurations, no matter how these
configurations have been generated. This is an important requirement to accurately predict
phase transitions. On the other hand, the quality of the metallic wave function on the 5
liquid configurations generated with this wave function is higher than on the remaining 10
liquid configurations generated in a LDABF run. A good trial
function should have a uniform quality throughout the entire proton configurational space
in order to provide an unbiased sampling.

A similar analysis has been performed at $r_s=1.31, 1.40$ and $1.55$ considering 5 liquid configurations generated during a CEIMC run at $T=2000K$ with the LDABF trial function. Results are displayed in Figures \ref{fig:rs1.31static}, \ref{fig:rs1.4static} and \ref{fig:rs1.55static} respectively. Since the metallic wave function is certainly not accurate at this density, we compare the IPP, LDA and LDABF wave functions only.
At all densities LDABF trial function provides lower energies (its VMC energy is very close the RQMC-IPP energy). Moreover its quality parameter is smaller than for the other wave functions because the backflow provides a considerable improvement already at the VMC level and the gap between the VMC and the RQMC energies is smaller than for the other cases. Also the quality parameter appears to be more uniform over the (small number of) configuration examined. Quality parameters of IPP and LDA trial functions are within error bars to each other except at $r_s=1.55$ where the IPP one is quite smaller, and close to the LDABF values, than the LDA values. This is caused by an larger energy gap between VMC and RQMC energies and a slower convergence in imaginary time for LDA trial function than for the other trial functions.

Energy dispersions are collected in Table \ref{table3}. We observe a large dependence of the dispersion (between 2 and 3 times) on the trial function and, for IPP and LDA, on the level of optimization of its bosonic part. As mentioned above, this fact would imply a large sensitivity of thermal properties to the type of trial function and on the QMC method exploited to obtain the BO energy surface. These conclusions are in agreement with our finding that the nature of the dissociation process changes from the VMC to the RQMC energy surface\cite{DelaneyPierleoniCeperley06}. In the LDABF case the difference in dispersion between VMC and RQMC energies corresponds to only few tens of degrees Kelvin, probably a negligible effect in most interesting cases. On the other hand VMC remains much more efficient in terms of computational resources required.

\subsection{Liquid-State Simulations}
After the validation of LDABF trial function of the previous section, we report here results for the liquid structure of hydrogen at the same densities. We have simulated systems of 54 protons. The TABC is performed here using twist sampling\cite{PierleoniDelaneyMoralesCeperleyHolzmann07} around the nodes of a 4x4x4 grid in twist space at each protonic step.
\begin{figure}[t]
\begin{center}
\psfig{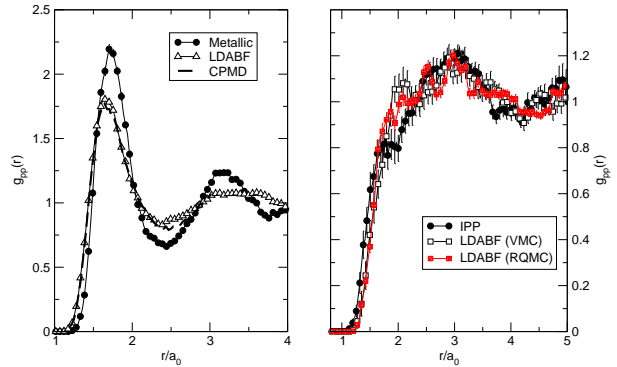}
\end{center}
\caption{Left panel: $r_s=1$, $T=1500K$, $N_p=54$. Proton-proton pair correlation functions as obtained with LDABF and metallic wave functions. Results are obtained with TABC by using twist sampling around a 4x4x4 grid. CPMD data from ref.\cite{KohanoffHansen96} are also represented by a thick dashed line. Right panel: $r_s=1.4$, $T=2000K$, $N_p=54$. Proton-proton pair correlation functions as obtained with IPP and LDABF wave functions. Results are obtained with TABC by using a 6x6x6 fixed grid (IPP) and by twist sampling around a 4x4x4 grid (LDABF).}
\label{fig:rs1DftbfvsMetGr}
\end{figure}
In the left panel of Figure \ref{fig:rs1DftbfvsMetGr} we report a comparison of proton-proton pair correlation functions $g_{pp}(r)$ at $r_s=1$ and $T=1500K$ as obtained from the metallic and LDABF trial functions at the VMC level. As expected from the results of the previous section, we observe considerably more structure with the metallic trial function than with the LDABF one, which indeed would correspond to having an effective lower temperature. On the same figure we report results from a CPMD simulation\cite{KohanoffHansen96} performed within the LDA approximation. The agreement between CPMD data and our present CEIMC data from LDABF trial function is striking and somehow unexpected. Indeed our representation of the electronic ground state is much more accurate than the simpler LDA one. Also the finite size effect in the CPMD calculation was addressed only partially by using only closed shells systems at the $\Gamma$ point. Nonetheless the observed agreement testify that the structure of the proton liquid is not very sensitive to details of the ground state representation, at least at such high density.
Finally, in the right panel of Figure \ref{fig:rs1DftbfvsMetGr} we report preliminary data for $g_{pp}(r)$ of 54 protons at $r_s=1.4$ and $T=2000K$. We compare IPP and LDABF trial functions at the VMC and RQMC levels. The statistical noise is still large but it seems that the overall behavior does not depend too much on the kind of trial functions, although small details could still be different. Note, however, that the liquid has little structure. More investigations of the influence of the trial function on the liquid structure is certainly needed, in particular, in the molecular dissociation region.

\section{Conclusions}\label{sec4}
We have reported important progress in CEIMC, an efficient and accurate method to perform \emph{ab-initio} simulations of condensed system with QMC energies. We have shown how the method performs in the case of hydrogen at high pressure, the simplest, but yet not understood, system. The new method allows us to cover a range of temperatures inaccessible to previous QMC methods for hydrogen, a range where most of the interesting physics of hydrogen occurs, including the melting of the molecular and proton crystals, the molecular dissociation both in the liquid and in the crystal and the metallization of the system.

A key ingredient in CEIMC is the trial function used to represent the electronic ground state. Even when a projection technique such as Reptation QMC is exploited to improve the bosonic part of the trial many-body wave function, its fermionic part, that is its nodal surface, is still playing a very crucial role in determining the electronic energies and therefore the overall thermal behavior of the system. In the present paper, we have reported a detailed investigation of these effects for hydrogen by comparing a number of different trial wave functions at two densities. We have shown as a fully analytical trial wave function, that is optimal in terms of computational efficiency in CEIMC, and which has been previously demonstrated to provide excellent accuracy for crystalline states, degrades as soon as some disorder is introduced in the protonic configurations. This result has been established by comparing with results for new trial functions obtained from a Slater determinant of IPP/LDA orbitals together with a two-body Jastrow correlation factor. A further backflow transformation of these orbitals has been introduced and characterized.
The new trial functions provide lower energies and more uniform overlap over a number of fixed representative configurations, which we use as an indication of the overall quality of the trial function. The most striking result on disordered configurations is that the LDABF energies at the VMC level are lower than the fully projected energies from the metallic trial function. This indicates that the improvement of performance comes mainly from the different nodal surfaces, while the bosonic part is responsible only for smaller improvements. The failure of the metallic wave function is most probably due to the presence of some degeneracy of its orbital structure around the Fermi surface which is removed by solving the instantaneous band structure. On the other hand, the use of complex wave functions and twist averaged boundary conditions in connection with the metallic trial function was expected to remove most of these degeneracies. A better understanding of this failure is desirable and deserves more investigation.

The difference in energies for different trial functions, or more precisely the dispersions of the energies from different wave functions, translates into a overall temperature factor at thermal equilibrium. The metallic trial function provides a dispersion which is roughly twice that of the corresponding dispersion from the LDABF trial function. Therefore the metallic $g_{pp}(r)$ at temperature $T$ should correspond to the LDABF $g_{pp}(r)$ at $\sim T/2$. This is indeed observed and the new  $g_{pp}(r)$'s from LDABF are in fair agreement with predictions of Car-Parrinello MD\cite{KohanoffHansen96}. This agreement remains somehow surprising since, beyond the different methods of sampling protonic configurational space, the electronic description in the two methods is quite different. We use LDA orbitals with a backflow transformation and a two body RPA Jastrow while in CPMD, only LDA orbitals are employed. Adding the backflow and the Jastrow, we obtain a fair gain of energy and, moreover, we can improve the bosonic part of the trial function by projecting in imaginary time. Further, we strongly reduce the finite size effects by averaging over the undetermined phase of the wave function, while CPMD calculations are performed at the $\Gamma$ point only for closed shell systems ($N_p=54$ and $162$). However the final agreement between the two methods indicates that the effects of these improvements on energy differences is only minor. On the other hand, it is well known in simple liquids that $g(r)$ is not very sensitive to changes of the interaction potential and this might explain the observed agreement.

At lower densities, employing IPP orbitals and RPA Jastrow, we have recently found\cite{DelaneyPierleoniCeperley06} a continuous molecular dissociation with density, at variance with CPMD which has predicted a first order molecular dissociation transition\cite{Scandolo03}. The reliability of IPP trial function was only tested on crystal structures and should be further investigated for disordered configurations along the lines shown here. This study is in progress. A recent BOMD study\cite{VorbergerTamblynMilitzerBonev07} within DFT/GGA has reported a continuous molecular dissociation in agreement with our findings. This agreement suggests that improving the trial functions from IPP to LDABF might change the details of the results but not the overall picture. This confirms that our present method can be most useful in condition where new interesting physics is happening, such as near a liquid-liquid phase transitions or a metallization transition.

\section*{Acknowledgments}
C.P. acknowledges financial support from MIUR-PRIN2005, D.M.C. and K.T.D. acknowledge support from DOE grant DE-FG52-06NA26170, M.A.M. acknowledges support from a DOE/NNSA-SSS graduate fellowship and M.H. acknowledges support from ACI ``Desorde et Interactions Coulombiennes" of French Ministry of Research and from CNISM, Italy. D.M.C. and M.H. acknowledge support from a UIUC-CNRS exchange program. Computer time was provided by NCSA, INCITE and CINECA.

\bibliographystyle{elsart-num}
\bibliography{rpmbt14}

\end{document}